# Cultural Rights and the Rights to Development in the Age of AI: Implications for Global Human Rights Governance

Alexander Kriebitz[1*], Caitlin Corrigan[2,] Aive Pevkur[3], Alberto Santos Ferro[4], Amanda Horzyk[5], Dirk Brand[6], Dohee Kim[7], Dodzi Koku Hattoh[8], Flavia Massucci[9], Gilles Fayad[10], Kamil Strzępek[11], Laud Ammah[2], Lavina Ramkissoon[12], Mariette Awad[13], Natalia Amasiadi[14], Nathan C. Walker[15], Nicole Manger[16] and Sophia Devlin[17]

*Main contact: a.kriebitz@lmu.de

[1] Ludwig Maximilian University of Munich; [2] Institute of Ethics in Artificial Intelligence, Technical University of Munich; [3] Tallinn University of Technology; [4] Santos Integrity International; [5] University of Edinburgh; [6] Stellenbosch University; [7] Changwon National University; [8] Bonn Sustainable AI Lab, Germany & University of Ghana; [9] Independent AI Ethics Researcher; [10] AI Commons; [11] Cardinal Stefan Wyszyński University in Warsaw; [12] African Union; [13] American University of Beirut; [14] University of Patras; [15] AI Ethics Lab, Rutgers University; [16] TUM Think Tank and Global Solutions Initiative (G7/20); [17] Ulster University.

**Abstract**

Cultural rights and the right to development are essential norms within the wider framework of international human rights law. However, recent technological advances in artificial intelligence (AI) and adjacent digital frontier technologies pose significant challenges to the protection and realization of these rights. This owes to the increasing influence of AI systems on the creation and depiction of cultural content, affect the use and distribution of the intellectual property of individuals and communities, and influence cultural participation and expression worldwide. In addition, the growing influence of AI thus risks exacerbating preexisting economic, social and digital divides and reinforcing inequities for marginalized communities. This dynamic situation challenges the existing interplay between cultural rights and the right to development, and raises questions about the integration of cultural and developmental considerations into emerging AI governance frameworks.

To address these challenges, the paper examines the impact of AI on both categories of rights. Conceptually, it analyzes the epistemic and normative limitations of AI with respect to cultural and developmental assumptions embedded in algorithmic design and deployment, but also individual and structural impacts of AI on both rights. On this basis, the paper identifies gaps and tensions in existing AI governance frameworks with respect to cultural rights and the right to development.

By situating cultural rights and the right to development within the broader landscape of AI and human rights, this paper contributes to the academic discourse on AI ethics, legal frameworks, and international human rights law. Finally, it outlines avenues for future research and policy development based on existing conversations in global AI governance.

# 1 Introduction

Artificial intelligence (AI) represents a paradigm shift in human history that increasingly permeates culture, development, and human rights. From media and arts to language and identity, AI influences the ideas, expressions, moral norms, religious concepts, and other cultural artefacts that shape our societies. This influence extends beyond technological novelty and reconfigures the very foundations of cultural participation, creative expression, and societal development on a global scale. Ultimately, generative AI, immersive virtual realities, and agentic AI have become forces transforming how culture is produced, experienced and embedded in socio-technical systems across different developmental contexts.

In this changing landscape, cultural rights and the right to development, articulated and deeply rooted in international law, face unprecedented challenges and opportunities.[1] The growing capabilities of AI, specifically agentic and generative AI, to reproduce, alter, manipulate, and represent cultural knowledge raise urgent anthropological, ethical, and legal questions that the concern the very interpretation of human rights. These developments demand a critical examination of how power, equity, and consent operate in AI ecosystems, as well as the mechanisms required to safeguard cultural diversity, minority identities, labor rights, and collective heritage, but also to guarantee meaningful remedies for individuals and groups adversely affected by AI.[2]

To address these emerging human rights challenge, the following paper aims to address the human rights impact of AI centering on cultural rights and the right to development, which have so far received little attention in human rights scholarship. In fact, most existing publications treat these issues rather as a side aspect of wider human rights impacts.[3]. Given their individual and collective dimensions, cultural rights and the right to development, however, represent a critical intersection between human rights conventions and principles articulated in international humanitarian law and the UN Charter.[4] They connect different dots of existing international law, but they also codify different legal interests of the international community.

To address this issue, the paper proceeds in three parts. The first section examines the existing role of cultural rights and the right to development in the existing normative context, including not only international legal norms, but also philosophical and cultural perspectives that shape the conversation around culture and development. The second section centres on the impact of AI on these rights, considering both immediate and structural impacts. Then, the paper examines the integration of cultural and developmental perspectives in exiting AI governance frameworks. On this basis, the paper offers recommendations for integrating cultural rights and the right to development into emerging discussions on AI and human rights.

---

[1] The following publication understands both types of rights as articulated in international legal documents, including Universal Declaration of Human Rights and the Covenant on Economic, Social and Cultural Rights, the UN Declaration of the Rights of Indigenious People as well as other regional human rights instruments. Further relevant international instruments addressing collective cultural rights and development include the UNESCO Convention on the Protection and Promotion of the Diversity of Cultural Expressions (2005), Hague Convention for the Protection of Cultural Property in the Event of Armed Conflict (1954), Council of Europe Framework Convention for the Protection of National Minorities (1995), and African Charter on Human and Peoples' Rights (1981). These codifications emphasize collective dimensions such as peoples' rights and cultural self-determination, complemented by legal scholarship and philosophical contributions that have influenced the interpretation of both rights.

[2] The academic community is increasingly aware of the wider systemic challenge AI poses to human rights as such. Many of the authors of the submission were involved in the Whitepaper „Promoting and advancing human rights in global AI ecosystems: The need for a comprehensive framework under international law." (Kriebitz, A., & Corrigan, C. C. (edit.). (2025). *Promoting and advancing human rights in global AI ecosystems: The need for a comprehensive framework under international law*. Ethics Lab at Rutgers University. https://aiethicslab.rutgers.edu/publications/promoting-and-advancing-human-rights-in-global-ai-ecosystems/

[3] See: Kriebitz, A., & Lütge, C. (2020). Artificial intelligence and human rights: a business ethical assessment. Business and Human Rights Journal, 5(1), 84-104. Risse, M. (2019). Human rights and artificial intelligence: An urgently needed agenda. Human Rights Quarterly, 41(1), 1-16.

[4] Borelli, S., & Lenzerini, F. (Eds.). (2012). Cultural heritage, cultural rights, cultural diversity: new developments in international law (Vol. 4). Martinus Nijhoff Publishers. Symonides, J. (1998). Cultural rights: a neglected category of human rights. *International Social Science Journal*, *50*(158). Sengupta, A. (2000). Realizing the right to development. *Development and Change*, *31*(3), 553-578.

## 2 The Normative Background of Cultural Rights and the Right to Development

Cultural rights and the right to development are grounded in a rich historical, legal, and intellectual tradition that expresses individual and collective human interests.[5] These rights are firmly embedded within the broader framework of international human rights law, reflecting centuries of efforts to secure human dignity, cultural diversity, and equitable progress for all peoples.[6] Recognizing their essential place in this normative context is crucial for ensuring that AI governance respects and promotes these rights as integral to global justice and sustainable development.

### 2.1. Cultural Rights within International Human Rights Law

The modern understanding of cultural rights traces back to Article 27 of the *Universal Declaration of Human Rights (UDHR)*, which recognizes the right of everyone to participate in cultural life and to share in scientific advancement and its benefits.[7] This principle was further enshrined in Article 15 of the *International Covenant on Economic, Social and Cultural Rights (ICESCR)*, affirming that cultural rights form an integral part of the universal human rights framework.[8]

That said, cultural rights encompass both individual and collective dimensions. Individuals have the right to express themselves through culture, language, and art without discrimination, while communities have the right to preserve, revitalize, and transmit their cultural heritage. This includes minority and indigenous groups whose languages, folklore, and traditions form part of humankind's collective heritage.[9]

In addition to cultural participation, these rights involve meaningful access to cultural institutions and resources, and the protection of artists and creators. They interconnect with other rights (such as the rights to education, work, and intellectual property) and are linked to the safeguarding of intangible cultural heritage, including traditional knowledge and expressions.[10]

Legal safeguards for cultural rights on the level of regional human rights conventions can be found in various instruments, from national constitutions to the *European Charter for Regional or Minority Languages, Art. 17 of the African Convention* for Human and Peoples' Rights and UNESCO conventions protecting cultural diversity.[11] The German constitutional context illustrates both the protection of artistic freedom and the potential conflicts it raises when reconciling cultural expression with the rights and dignity of others.[12] Cultural freedom is therefore not unlimited; it must coexist with principles of equality, non-discrimination, and respect for human dignity.[13] In addition, cultural rights such as artistic freedom also have contextual dimensions such as irony or sarcasm that are context specific and subject to wider societal preferences and customs.

---

[5] Nussbaum, M. (2007). Human rights and human capabilities. Harv. Hum. Rts. J., 20, 21.
[6] Buergenthal, T. (2006). The evolving international human rights system. The American Journal of International Law, 100(4), 783-807. D'Amato, A. (2017). The concept of human rights in international law. In International Law of Human Rights (pp. 21-70). Routledge.
[7] Langford, M. (Ed.). (2013). Global justice, state duties: the extraterritorial scope of economic, social, and cultural rights in international law. Cambridge University Press.
[8] OHCHR. (2025). International Covenant on Economic, Social and Cultural Rights. (1966). https://www.ohchr.org/en/instruments-mechanisms/instruments/international-covenant-economic-social-and-cultural-rights
[9] Borelli, S., & Lenzerini, F. (Eds.). (2012). Cultural heritage, cultural rights, cultural diversity: new developments in international law (Vol. 4). Martinus Nijhoff Publishers.
[10] Francioni, F. (2011). The human dimension of international cultural heritage law: an introduction. European Journal of International Law, 22(1), 9-16. Kalantry, S., Getgen, J. E., & Koh, S. A. (2010). Enhancing enforcement of economic, social, and cultural rights using indicators: A focus on the right to education in the ICESCR. Human Rights Quarterly, 32(2), 253-310.
[11] African Union (1981). African Charter on Human and People's Rights. Retrieved from: https://au.int/en/treaties/african-charter-human-and-peoples-rights

## 2.2 Right to Development as an Expression of Individual and Collective Rights

The right to development, recognized by the *UN Declaration on the Right to Development* (1986), establishes that all peoples and individuals are entitled to participate in, contribute to, and enjoy economic, social, cultural, and political development. It implies that development must aim to fully realize human rights and fundamental freedoms, ensuring the well-being of entire populations.[14] The right to development thus bridges individual and collective rights, linking participation in cultural and economic life with broader agendas of self-determination and justice on a global level characterized by economic, developmental and social inequalities.

Hence, the right to development includes principles of equity, participation, and international cooperation, while imposing obligations on states. It also reflects the duty of states to pursue development in ways that reduce inequalities, respect sovereignty over natural resources, and ensure the fair distribution of benefits.[15]

Historically, the right to development has received strong support in the Global Majority.[16] This owes to the fact that violations of this right often affect marginalized and indigenous communities, who traditionally face exclusion from the benefits of technological and economic progress. Subsequently, the right to development has been enshrined in the African Convention on Human and Peoples' Rights in Art. 22.[17] In addition, the Universal Declaration of the Rights of Indigenous People makes explicit references to the right of development. Article 23, for instance, reaffirms the "right to determine and develop priorities and strategies for exercising their right to development."[18]

Relatedly, normative considerations with respect to development are based on the premise that societies face different starting conditions and capabilities for realizing international legal norms.[19] This notion

---

Mancini, S., & De Witte, B. (2008). Language rights as cultural rights: a European perspective. *International Studies in Human Rights*, *95*, 247.

[12] Maroz, R. (2016). The Freedom of Artistic Expression in the Jurisprudence of the United States Supreme Court and Federal Constitutional Court of Germany: A Comparative Analysis. *Cardozo Arts & Ent. LJ*, *35*, 341.

[13] See: Federal Constitutional Court (Bundesverfassungsgericht). (2007, June 13). Order of the First Senate of 13 June 2007, 1 BvR 1783/05. Retrieved from https://www.bundesverfassungsgericht.de/SharedDocs/Entscheidungen/EN/2007/06/rs20070613_1bvr178305en.html

[14] United Nations General Assembly. (1986). Declaration on the Right to Development (A/RES/41/128). Retrieved from https://www.refworld.org/legal/resolution/unga/1986/en/15508

[15] Barsh, R. L. (1991). The right to development as a human right: results of the global consultation. *Human Rights Quarterly*, *13*(3), 322-338.

[16] The following publication uses the term Global Majority for accounting for underpriviledged voices in the global discourse, due to developmental, geographical or other historical factors. See: Stanford Institute for Human-Centered Artificial Intelligence. (2023). *Moving beyond the term "Global South" in AI ethics and policy*. Stanford University. https://hai.stanford.edu/policy/moving-beyond-the-term-global-south-in-ai-ethics-and-policy

17 Kamga, S. A. D. (2011). The right to development in the African human rights system: The Endorois case. De Jure, 44(2), 381-391. Ngang, C. C. (2018). Towards a right-to-development governance in Africa. Journal of Human Rights, 17(1), 107-122. Uwazuruike, A. (2020). *Human rights under the African charter*. Cham: Springer International Publishing.

[18] United Nations Declaration on the Rights of Indigenous Peoples. https://www.un.org/development/desa/Indigenouspeoples/wp-content/uploads/sites/19/2018/11/UNDRIP_E_web.pdf

[19] Nussbaum, M. (2007). Human rights and human capabilities. Harv. Hum. Rts. J., 20, 21.

has been particularly discussed in the context of progressive realization of economic, social and cultural rights. States with a lower level of development cannot meaningfully realize economic, social, and cultural rights, for instance, when it comes to access to health or social security, to the same extent as countries of the Global North. The idea of a capability approach that ties duties under international law to actors' capabilities in realizing human rights has therefore emerged as an essential lens within international law for addressing such inherent asymmetries.[20] A major example for including development constraints within international legal frameworks is the Kyoto Protocol. Particularly noteworthy within this protocol is here the principle of 'Common But Differentiated Responsibilities and Respective Capabilities' (CBDR-RC), which holds all states accountable for addressing human-driven climate change, it also addresses asymmetries in the realization of climate policies.[21]

Hence, the acknowledgement of differences in capabilities to realize norms underpins contemporary debates about business ethics, sustainable development and social responsibility within technological innovation, by acknowledging the limits and constraints of actors when realizing norms.[22] Conversely, scholarship has highlighted the moral duty of mutual support within the international community, emphasizing the need to give special consideration to countries and societies with weaker starting positions.[23] This duty reflects broader normative aims, such as the resolution of "international problems of an economic, social, cultural, or humanitarian character," which have already been identified as goals of the United Nations.[24]

## 2.3 Cultural Rights and the Right to Development as Norms within Human Rights Law

Cultural rights and the right to development cannot be understood in isolation, particularly when examining both categories of rights from a global perspective. Instead, both are embedded within the broader normative framework of international human rights law and derive their meaning from the interplay between universal principles, state obligations, and evolving interpretive practices.[25] While they share a common orientation towards human dignity as acknowledged by the OHCHR[26], collective participation, and self-determination, their scope and acceptance within international law differ significantly.

Cultural rights enjoy a well-established position under the Universal Declaration of Human Rights (Article 27) and the International Covenant on Economic, Social and Cultural Rights (Article 15), which recognize the right of everyone to participate in cultural life and to share in scientific advancement.[27] The right to development, by contrast, has historically been more contested.[28] Although codified in the 1986 UN Declaration on the Right to Development and reaffirmed at the 1993 World Conference on

---

[20] Sen, A. (1990). Development as capability expansion. Human development and the international development strategy for the 1990s, 1(1), 41-58.
[21] Matsui, Y. (2004). The principle of "common but differentiated responsibilities". In International law and sustainable development (pp. 73-96). Brill Nijhoff.
[22] See: Enderle, G. (2021). Corporate responsibility for wealth creation and human rights. Cambridge University Press.
[23] Martin, R. (2015). Rawls on international economic justice in the law of peoples. Journal of Business Ethics, 127(4), 743-759. Villaroman, N. (2010). The right to development: Exploring the legal basis of a supernorm. *Florida Journal of International Law, 22*, 299–332.
[24] See: United Nations. (1945). Charter of the United Nations. https://www.un.org/en/about-us/un-charter/full-text
[25] Radin, M. L. (1993). The Right to Development as a Mechanism for Group Autonomy: Protection of Tibetan Cultural Rights. *Wash. L. Rev., 68*, 695. Symonides, J. (1998). Cultural rights: a neglected category of human rights. *International Social Science Journal, 50*(158).
[26] Office of the United Nations High Commissioner for Human Rights. (2025). OHCHR and the right to development. United Nations. https://www.ohchr.org/en/development#:~:text=Over%20thirty%20years%20ago%2C%20the%20Declaration%20on%20the,to%20participate%20fully%20and%20freely%20in%20vital%20decisions.
[27] Riedel, E., Giacca, G., & Golay, C. (Eds.). (2014). *Economic, social, and cultural rights in international law: contemporary issues and challenges*. OUP Oxford.
[28] Marks, S. (2004). The human right to development: between rhetoric and reality. *Harv. Hum. Rts. J., 17*, 137.

Human Rights in Vienna, its legal status remains debated, with some Western states reluctant to recognize it as a binding right under customary international law.[29]

Despite these differences, both rights converge in their aim to advance human well-being and equitable participation in the cultural, social, and economic life of humanity.[30] They are closely connected to the principle of self-determination, as articulated in Articles 1, 55, and 56 of the UN Charter, which emphasizes international cooperation to promote social progress and higher standards of living.[31] Cultural rights provide the normative foundation for expressing collective identity, preserving cultural diversity, and protecting the heritage of indigenous and minority communities, but also exemplify different approaches on economic, social and cultural development. This is firmly rooted in the broader principle of equality outlined in the UN Charter, which enshrines equality of all individuals regardless of race, sex, language, or religion, which constitutes a dimension deeply intertwined with our understanding of culture. Therefore, cultural rights are inherently embedded within UN Charter, reflecting the cultural dimensionality of equal dignity and rights of all human beings.[32]

At the same time, neither right operates without constraints. The pursuit of development has at times been associated with infringements on cultural or individual rights, through forced assimilation, displacement of minority populations, or environmental degradation caused by large-scale industrial projects.[33] Cultural rights can also intersect and occasionally conflict with intellectual property law, freedom of expression, or individual privacy, especially in contexts involving indigenous knowledge, artistic production, or digital media. These tensions underscore the need for a balanced interpretation that integrates cultural and developmental goals without undermining other human rights and the principle of self-determination of nations.

International Humanitarian Law (IHL) and the Genocide Convention add further dimensions by recognizing the protection of cultural identity and memory even during conflict, including the prohibition of acts amounting to cultural genocide or the systematic erasure of cultural heritage.[34] These provisions highlight that the preservation of culture is not only a matter of human rights but also of international peace and security. Consequently, research in the transitional justice literature increasingly highlights the relevance of cultural rights as an important dimension of positive peace, while also pointing to growing awareness of the impact of policies on marginalized cultures.[35]

In sum, cultural rights and the right to development form part of a broader, interdependent web of international legal norms that ultimately aim at human flourishing.[36] They connect not only to international human rights law, but also need to be seen in a wider international legal context, including considerations pertaining to human dignity, but also the collective right to self-determination. This connection is explicitly addressed in international human rights law, particularly the UN Declaration on the Rights of Indigenous Peoples.[37] In conclusion, the effective realization depends on maintaining a dynamic balance between collective advancement, individual freedoms, and respect for diversity. This understanding is crucial in the emerging context of AI, where technology increasingly influences cultural expression, identity formation, and equitable access to development.

---

[29] Jha, S. (2012). A critique of right to development. Journal of Politics & Governance, 1(4), 17-22.
[30] See: United Nations General Assembly. (2024). The right to development (A/RES/59/185). https://digitallibrary.un.org/record/4071786?v=pdf; United Nations General Assembly. (1998). Right to development (A/RES/41/128). https://digitallibrary.un.org/record/265884?ln=en&v=pdf
[31] Office of the United Nations High Commissioner for Human Rights. (1986). Declaration on the right to development. United Nations. https://www.ohchr.org/en/instruments-mechanisms/instruments/declaration-right-development
[32] See: United Nations. (1945). Charter of the United Nations. https://www.un.org/en/about-us/un-charter/full-text
[33] Arimoto, Y., & Lee, C. (2021). Domestic industrialization under colonization: evidence from Korea, 1932–1940. European Review of Economic History, 25(2), 379-403.
[34] Morsink, J. (1999). Cultural genocide, the Universal Declaration, and minority rights. Human Rights Quarterly, 21(4), 1009-1060. Mako, S. (2012). Cultural genocide and key international instruments: Framing the indigenous experience. International Journal on Minority and Group Rights, 19(2), 175-194. Vrdoljak, A. F. (2009). Cultural heritage in human rights and humanitarian law. International Human Rights and Humanitarian Law, 250-302.
[35] Luoma, C. (2021). Closing the cultural rights gap in transitional justice: Developments from Canada's National Inquiry into Missing and Murdered Indigenous Women and Girls. Netherlands Quarterly of Human Rights, 39(1), 30-52. Mutua, M. (2015). What is the future of transitional justice?. *International Journal of Transitional Justice*, *9*(1), 1-9.
[36] Nussbaum, M. C. (1997). Capabilities and human rights. Fordham L. Rev., 66, 273.

## 3 The Comprehensive Impact of AI on Cultural Rights and the Right to Development

The impact of AI on human rights, particularly cultural rights and the right to development, manifests through multiple, interrelated channels. They can arise directly from AI systems and their outputs, indirectly from the social and economic infrastructures that sustain AI ecosystems, or more diffusely from their influence on cultural and developmental processes across societies. For analytical clarity, these can be distinguished as impacts *created by AI* or *linked to AI*.

While a comprehensive assessment of all these impacts is not yet possible, the following section identifies analytical lenses to examine how AI reshapes the structure, expression, and enjoyment of cultural rights, and how it conditions the realization of the right to development.

### 3.1 The Human Rights Relevance of AI

While AI can be used for different purposes, it cannot be viewed as a neutral technology. It encompasses processes of data collection, algorithmic training, automated decision-making, and processes of human-machine interaction that embed social values, economic priorities, political preferences and anthropological assumptions.[38] At each stage of the AI lifecycle, from data acquisition to the eventual decommissioning of AI systems, human rights considerations might involve complex questions relating to human dignity, human autonomy, non-discrimination, the right to remedy, and their contextualization in applied contexts such as health, jurisprudence, defence, law enforcement, education or media.[39] Moreover, the design of AI systems often involves norm derogation and the application of theoretical approaches, creating substantial opportunities for variations in the discretion of private and public entities involved in the AI lifecycle when making decisions that affect human rights.[40]

From a human rights perspective, AI may amplify existing inequalities or create new forms of exclusion. Algorithmic bias can reproduce stereotypes or reinforce structural discrimination; opaque decision-making processes can erode individual autonomy and procedural fairness, and intensive data extraction can threaten privacy and informational self-determination.[41] The cumulative effect of these risks is not merely technical but also sociotechnical, affecting the protection, realization, and enforcement of human rights, particularly as AI increasingly involves itself in areas critical to their enjoyment. Examples of

---

this include AI systems used in triage contexts for health applications, or AI systems deployed by courts for evidence evaluation or psychological assessments.[42]

While these general human rights risks are widely recognized in emergent AI regulation and policy frameworks such as the EU AI Act, their cultural and developmental dimensions have received less attention so far.[43] Yet every AI system carries implicit cultural representations in its dataset, design, and interfaces that influence how cultural identities are portrayed, preserved, or marginalized.[44] Likewise, the global distribution of AI innovation reflects uneven developmental patterns, thus limiting equitable participation in the emerging digital economy. A major concern involves transnational harms arising from AI and the challenges of ensuring accountability across a global AI value chain, which includes diverse actors responsible for data procurement, model development, deployment, and eventually the discontinuation of the system.[45] This diffusion of responsibilities has been addressed early on by human rights scholarship, but is pertinent for context where the implications of human rights are themselves not straightforward.

### 3.2 The Epistemic Limitations of AI in Cultural and Developmental Contexts

AI systems are fundamentally data-driven. Reasoning is grounded in statistical correlations derived from human-provided data, and what is absent from the data remains invisible to the system.[46] AI depends, therefore, on the implicit and explicit assumptions of developers, who make relevant design choices, including morally relevant designs such as human-machine interfaces or the selection and quantification of variables. This set-up poses risks, particularly when involving complex terms related to culture of development.

This limitation has profound implications for concepts—and some might argue social constructs—such as art, culture, moral values, religion, and development, that are interpretive, context-bound and difficult to quantify or operationalize.[47] In addition, language plays a critical role in expressing these concepts, as it both reflects and constructs cultural identity and worldviews. Terms such as "sustainability" (Nachhaltigkeit, durabilité, jätkusuutlikkus, inbhuanaitheacht, الاستدامة) and "responsibility" (Verantwortung, responsabilité, vastutus, freagracht, المسؤولية) carry distinct normative and historical connotations in German, French, Estonian, Irish, and Arabic contexts, respectively. Similar complexities arise in the context of the depiction of religions and moral traditions through AI, particularly LLMs.[48] Philosophical and legal discussions highlight the inherent challenges in standardizing such terms, by pointing to the diversity of moral traditions like Ubuntu, Harambee or the ethics of the Ga people in Ghana or by challenging established concepts of regional identities and

---

[42] Tahernejad, A., Sahebi, A., Abadi, A. S. S., & Safari, M. (2024). Application of artificial intelligence in triage in emergencies and disasters: a systematic review. BMC Public Health, 24(1), 3203. Buschmeyer, K., Hatfield, S., & Zenner, J. (2023). Psychological assessment of AI-based decision support systems: tool development and expected benefits. Frontiers in Artificial Intelligence, 6, 1249322.

[43] see here for the EU AI Act: Regulation (EU) 2024/1689 of the European Parliament and of the Council of 13 June 2024 laying down harmonised rules on artificial intelligence. Official Journal of the European Union. (14.7.2024). Regulation (EU) 2024/1689 of the European Parliament and of the Council. (2024). https://eur-lex.europa.eu/eli/reg/2024/1689/oj/eng

[44] Prabhakaran, V., Qadri, R., & Hutchinson, B. (2022). Cultural incongruencies in artificial intelligence. arXiv preprint arXiv:2211.13069.

[45] Hohma, E. (2024). On the elements and implications of accountability for AI providers. In The Elgar Companion to Applied AI Ethics (pp. 13-36). Edward Elgar Publishing.

[46] Williams, B. A., Brooks, C. F., & Shmargad, Y. (2018). How algorithms discriminate based on data they lack: Challenges, solutions, and policy implications. Journal of Information Policy, 8, 78-115.

[47] See: Wallace, B. C. (2015). Computational irony: A survey and new perspectives. Artificial intelligence review, 43(4), 467-483. Kanuck, S. (2019). Humor, ethics, and dignity: Being human in the age of artificial intelligence. Ethics & International Affairs, 33(1), 3-12.

[48] Tsuria, R., & Tsuria, Y. (2024). Artificial intelligence's understanding of religion: investigating the moralistic approaches presented by generative artificial intelligence tools. Religions, 15(3), 375.

Cultural Rights and the Rights to Development in the Age of AI:
Implications for Global Human Rights Governance

cultural macro-regions.[49] Ultimately, culture, moral, religious and developmental assumptions and concepts are subject to change. For example, evolving meanings of words, slang, and societally accepted expressions may not be promptly captured. In a similar vein, developmental assumptions can be asymmetric due to the lack of up-to-date data, including fundamental economic indicators.[50] This data latency can lead to outdated or skewed AI outputs, especially affecting less digitally represented regions and communities, while posing inherent limitations to the digital representation of development, culture and related concepts.

Consequently, algorithmic systems that need to operationalize assumptions in respect to culture and development often reproduce prevailing behavioural, cultural and developmental assumptions embedded in training datasets. Thus, AI systems operate within the perspectives of their developers and operators, thereby perpetuating human biases and limiting cultural sensitivity, but also leading to a loss of cultural nuance and sensitivity.

Apart from the limitations of AI as a data-processing system, the use of AI raises a host of anthropological and ethical questions about what constitutes creativity or authorship in cultural production. They relate directly to human dignity as articulated in Article 1 of the UDHR, which emphasizes the intrinsic worth and individuality of every person and acknowledges the complexity of human life. The question of the comparison between human and machine intelligence in the cultural space is therefore not only categorical, but also concerns the individuals whose lived experiences, cultural expressions, languages and moral values are more embedded within AI systems.[51] The conceptualization of AI as a data-driven and statistical tool affects interpersonal and collective relations. The use of chatbots and synthetic media in communication may alter cultural norms of interaction, empathy, and authority.[52] Thus, epistemic limits intersect with social processes of cultural transmission and belonging, influencing how communities sustain their traditions and values across generations. This perspective relates to ongoing conversations in philosophy and cultural studies on the intrinsic relationship between individuals and their cultural or historical contexts, which also encompass the formation of identity, which is increasingly shaped in digital environments and by algorithmic structures[53]. Hence, the epistemic limitations on AI can touch upon different human rights contexts, while reinforcing perspectives on alienation, representation and belonging in societies that have cultural dimensions.

Taken together, the epistemic limitations of gaps raise distinct issues inherent to the use of AI in cultural and developmental contexts. A major aspect resides here in the methodology of AI as a statistical tool for evaluating data and the difficulties of operationalization and quantifying concepts such as cultural, values or development in the AI context. Ultimately, the epistemic limitations of AI present a wider a wider challenge for the realization of human dignity and self-determination in the context of AI use in culturally relevant settings.

---

[49] Mignolo, W. D. (2009). The Idea of Latin America. John Wiley & Sons. Ammah, L. N. A., Lütge, C., Kriebitz, A., & Ramkissoon, L. (2024). AI4people− an ethical framework for a good AI society: the Ghana (Ga) perspective. Journal of Information, Communication and Ethics in Society, 22(4), 453-465.
[50] Fang, S., & Han, Z. (2025). On the nascency of ChatGPT in foreign language teaching and learning. Annual Review of Applied Linguistics, 45, 148–178. doi:10.1017/S026719052510010X
[51] Nakadai, R., Nakawake, Y., & Shibasaki, S. (2023). AI language tools risk scientific diversity and innovation. Nature Human Behaviour, 7(11), 1804-1805.
[52] Krügel, S., Ostermaier, A., & Uhl, M. (2023). Algorithms as partners in crime: A lesson in ethics by design. Computers in Human Behavior, 138, 107483. Krügel, S., Ostermaier, A., & Uhl, M. (2023). ChatGPT's inconsistent moral advice influences users' judgment. Scientific Reports, 13(1), 4569. Srinivasan, R., & González, B. S. M. (2022). The role of empathy for artificial intelligence accountability. Journal of Responsible Technology, 9, 100021.
[53] Nussbaum, M. C. (2007). Frontiers of justice: Disability, nationality, species membership. In Frontiers of justice. Harvard University Press. Nussbaum, M. C. (2019). The cosmopolitan tradition: A noble but flawed ideal. Belknap Press.

### 3.3 AI Development as an Expression of Cultural Rights and the Right to Development

In spite of the limitations of AI, the development of the technology can also act as an expression of cultural rights and the right to development. Technological innovation is, in itself, a manifestation of human creativity, collective knowledge and scientific progress.[54] Human creativity and scientific progress are explicitly protected by the Universal Declaration of Human Rights as a contextualization of an anthropology based on human autonomy and creativity. The design and application of AI systems reflect cultural diversity, value systems, and societal priorities, whether in creating generative art, simulating traditional music, preserving endangered languages through translation models or creating new forms of religious expression.[55] Given the inherent openness of the term "culture", AI itself can be seen as an expression of culture, including AI-generated memes or its use to create art itself, for instance, in the context of virtual realities, robotics, or computer games.[56] That said, the development of AI, and in a wider sense innovation, is itself connected to different human rights, including artistic freedom, freedom of opinion and expression, or economic and entrepreneurial freedom.[57]

From a developmental perspective, AI holds transformative potential. If properly governed, it can contribute to education, cultural participation, and equitable access to knowledge. It can support the translation and dissemination of cultural materials across minority languages, thus enhancing intercultural understanding. The digitization of texts and books has already contributed to a better understanding of history, but also the preservation of cultural heritage. Projects such as AI-assisted linguistic preservation for endangered languages exemplify how technology can advance both cultural rights and developmental goals.[58] Given its relevance as a driver for socio-economic development, for instance, when considering the creation of new economies, but also giving individuals access to information, education and health, AI and adjacent digital technologies can be significant tools for the progressive realization of the right to development and closely related economic, social, and cultural rights.

However, the right to development also imposes obligations on state agencies: access to AI innovations and their benefits must be equitable, inclusive, and environmentally sustainable. The digital divide between the Global North and the Global Majority manifested in access to computing power, skilled labor, and data infrastructure continues to hinder the realization of this right.[59] In fact, the uneven development of AI can recreate new inequalities along developmental lines.[60] Moreover, societies that implement AI earlier than others might have important first-mover advantages over societies that lack the infrastructural environments to implement AI. Structural exclusion of individuals and nations from this, for instance, through restrictions in international trade or through other barriers of access to AI

---

[54] Bloomfield, B. P. (2018). The culture of artificial intelligence. In The question of artificial intelligence (pp. 59-105). Routledge.
[55] David, O. O., & Oyekunle, M. O. (2025). AI and Colour Symbolism in Nigerian Religions: A Religio-Ethical Study. African Journal of Religious and Theological Studies, 5(1), 105-124.
[56] Olena, P., Iryna, V., Nataliia, K., & Volodymyr, F. E. D. (2020). Memes as the phenomenon of modern digital culture. Wisdom, (2 (15)), 45-55. Ivcevic, Z., & Grandinetti, M. (2024). Artificial intelligence as a tool for creativity. Journal of Creativity, 34(2), 100079.
[57] Betta, M., Jones, R., & Latham, J. (2010). Entrepreneurship and the innovative self: A Schumpeterian reflection. International Journal of Entrepreneurial Behavior & Research, 16(3), 229-244.
[58] Nanduri, D. K., & Bonsignore, E. M. (2023). Revitalizing Endangered Languages: AI-powered language learning as a catalyst for language appreciation. arXiv preprint arXiv:2304.09394.
[59] African Commission on Human and Peoples' Rights. (2021). Resolution on the need to undertake a study on the impact of artificial intelligence, robotics, and other new and emerging technologies on human and peoples' rights in Africa (ACHPR/Res. 473 (2021)). Retrieved from https://achpr.au.int/en/adopted-resolutions/473-resolution-need- undertake-study-human-and-peoples-rights-and-art.
[60] Božić, V. (2023). Artifical intelligence as the reason and the solution of digital divide. Language Education and Technology, 3(2).

Cultural Rights and the Rights to Development in the Age of AI:
Implications for Global Human Rights Governanceresources such as legislation, can thereby be regarded as an obstacle to the realization of human rights in the context of AI.[61]

The right to development, but also cultural rights, therefore, implies that establishing equal opportunities for individuals and nations to participate in the development and use of AI, particularly in areas relevant to these rights, is necessary from a human rights perspective. Owing to the different starting conditions of nations in the age of AI, this also implies that nations have the freedom to choose how to engage in AI governance and to pursue specific development strategies. This aligns with existing conversations on the concretization of existing economic, social and cultural rights to basic public goods, for instance the right to access to electricity or the right to access to internet.[62] A major question is here, whether certain AI models will become so elementary that structural non-inclusion amounts to the violation of human rights. Such perspectives could be derived from interpretation of individual human rights, for instance the right to information (UDHR Art. 19) or the right "to share in scientific advancements and benefits" (UDHR Art. 27) and collective rights, for instance the Declaration of the Right to Development. The true extent of such translational rights, that seem to indicate a minimum standard for access to or participation in basic AI systems, however, requires careful consideration, given the critique in human rights literature on the creation of new human rights and its impact on existing human rights law[63].

Ultimately, the right to development and the understanding of AI as an expression of cultural rights are, while important, not unconditional. In fact, the alignment of AI with cultural rights and the right to development requires a higher consideration of potential and actual adverse impacts created by AI, impacts linked to AI, as well as the relevance of cultural and developmental assumptions within different stages of the AI lifecycle.

### 3.4 Impacts Created by AI

Impacts *created by AI* refer to the immediate cultural and ethical consequences of algorithmic processes and their outputs. These include biased representations, cultural misappropriation, and the erosion of cultural autonomy through automated reproduction and circulation of symbolic materials.

At the *individual level*, AI systems can undermine artistic and intellectual integrity when creative works, voices, or likenesses are used without consent or compensation.[64] Such practices infringe upon moral and material interests protected under Article 27 of the UDHR.[65] When an artist's painting or composition is ingested into a dataset without credit or consent, or when a generative model meaningfully imitates an individual's style, identity, or voice, it undermines both cultural pride and personal rights to authorship and expression.[66] Similar concerns exist for performers, musicians, and writers whose creative identities are algorithmically replicated.[67]

At the *collective level*, the consequences extend to cultural groups and communities. Indigenous peoples and local communities often rely on the commercialization of cultural symbols, folklore, and traditional crafts as essential sources of income and avenues for sustaining culturally specific lifestyles. When generative AI reproduces traditional patterns or motifs without acknowledgment, meaningful engagement, or financial return, it can deprive these communities of both economic resources and cultural recognition.[68] This not only violates principles of cultural rights but also undermines indigenous peoples' broader rights to self-determination and to the maintenance of distinct cultural, social, and economic structures, as recognized by the UN Declaration on the Rights of Indigenous Peoples (UNDRIP).[69]

---

[61] See for earlier precedent cases: Bunn, I. D. (1999). The right to development: Implications for international economic law. Am. U. Int'l L. Rev., 15, 1425.
[62] Tully, S. (2014). A human right to access the internet? Problems and prospects. Human Rights Law Review, 14(2), 175-195. Burgess, R., Greenstone, M., Ryan, N., & Sudarshan, A. (2020). The consequences of treating electricity as a right. *Journal of Economic Perspectives*, *34*(1), 145-169.
[63] Skepys, B. (2012). Is there a human right to the internet. J. Pol. & L., 5, 15.
[64] Geiger, C., & Iaia, V. (2024). The forgotten creator: Towards a statutory remuneration right for machine learning of generative AI. Computer law & security review, 52, 105925.
[65] Geiger, C. (2024). Elaborating a human rights-friendly copyright framework for generative AI. IIC-International Review of Intellectual Property and Competition Law, 55(7), 1129-1165.
[66] Genelza, G. G. (2024). A systematic literature review on AI voice cloning generator: A game-changer or a threat?. Journal of Emerging Technologies, 4(2), 54-61.
[67] Dugeri, M. (2024). The Cannibalization of Culture: Generative AI and the Appropriation of Indigenous African Musical Works. J. Intell. Prop. & Info. Tech. L., 4, 17.

The collective dimension further extends to the right to freedom of thought, conscience, religion, or belief, which protects both individuals and communities in expressing and practicing shared worldviews. When AI systems position themselves as authoritative sources of information, particularly in moral, cultural, or religious domains, they risk distorting pluralism by privileging certain interpretations or belief systems. This is particularly relevant against the backdrop of research in behavioural research, which suggests that AI is influencing human judgment, speech patterns in language, and normative positions.[70] Thus, ensuring that AI-driven knowledge reflects a plurality of opinions and traditions is essential to maintaining cultural diversity and protecting the collective autonomy of belief communities.[71]

Moreover, the informational self-determination of groups becomes increasingly important in the digital age. Communities have a legitimate interest to deciding how their cultural data, traditions, or collective knowledge are represented in or by AI systems, particularly if these systems are accessible to decision-makers or the wider public. This aligns with broader debates on information and data sovereignty and on collective data governance, where communities demand participation in decisions that affect their cultural representation, proliferation, and identity.

These collective rights take on heightened importance in political contexts marked by assimilation pressures or the contestation of cultural identities. Minority populations in several regions remain vulnerable to policies of cultural homogenization. Examples include the erosion of linguistic autonomy among Russian-speaking minorities in the Baltic States[72], the marginalization of Ukrainian-speaking communities and Crimean Tatars under Russian occupation[73], and the social stigmatization of cultural identities that may be mirrored or reinforced by linguistically or culturally biased AI models. AI thus reflects and can amplify structural inequalities already present in political and societal contexts.

---

[68] Tapu, I. F., & Fa'agau, T. K. I. (2022). A new age indigenous instrument: Artificial intelligence & its potential for (de) colonialized data. Harv. CR-CLL Rev., 57, 715.

[69] Importantly, the UNESCO Recommendations on Ethics of Artificial Intelligence have addressed this issues in multiple ways. See also: United Nations. (2007). United Nations Declaration on the Rights of Indigenous Peoples. https://www.un.org/development/desa/Indigenouspeoples/wp-content/uploads/sites/19/2018/11/UNDRIP_E_web.pdf

[70] Krügel, S., Ostermaier, A., & Uhl, M. (2023). ChatGPT's inconsistent moral advice influences users' judgment. Székely, É., Miniota, J., & Hejná, M. (2025). Will AI shape the way we speak? The emerging sociolinguistic influence of synthetic voices. arXiv. https://arxiv.org/abs/2504.10650

[71] Abrar, A., Oeshy, N. T., Kabir, M., & Ananiadou, S. (2025). Religious bias landscape in language and text-to-image models: Analysis, detection, and debiasing strategies. AI & SOCIETY, 1-27.

[72] Ubarevičienė, R., & Burneika, D. (2025). The Russian minority in the Baltic capitals: Examining marginalisation in the context of urban dynamics. In Urban Marginality, Racialisation, Interdependence (pp. 145-166). Routledge. Sarkar, B., Dutta, A. (2024). The Lingering Question of Russian Minorities in Baltic Nation-States. In: Sarkar, B. (eds) The Baltics in a Changing Europe. Palgrave Macmillan, Singapore. https://doi.org/10.1007/978-981-97-5890-6_16

[73] Office of the United Nations High Commissioner for Human Rights. (2024, March 19). Human rights situation during the Russian occupation of territory of Ukraine and its aftermath. United Nations. https://www.ohchr.org/en/documents/country-reports/human-rights-situation-during-russian-occupation-territory-ukraine-and

## 3.5 Impacts Linked to AI

Impacts *linked to AI* arise from the broader sociotechnical structures surrounding algorithmic systems. These systems mediate access to cultural representation, language, and expression, thereby influencing the societal infrastructure of human rights enjoyment.[74] Impacts linked to AI are particularly relevant concerning the equitable realization of cultural rights and the right of development in "online" and "offline" spaces alike.[75] Moreover, impacts linked to AI need to be understood in the light of cumulative human rights impacts, where the interplay between several actors in an ecosystem gives rise to impact at a systemic level.[76]

Generative and conversational AI systems have become dominant channels for communication and content creation. However, they often reproduce linguistic hierarchies, providing higher quality interactions in globally dominant languages such as English or Mandarin, while degrading performance for smaller or under-resourced languages such as Basque, Corsican, Gaelic, Navajo, Nahuatl or Sorbian.[77] This technological bias risks accelerating the decline of linguistic or cultural diversity, a phenomenon historically observed during earlier technological transitions and modernization efforts, particularly in the wake of globalization, leading to the marginalization of cultural minorities and indigenous peoples.[78] In such cases, it is not necessarily one particular system, which is responsible for such adverse impact, but rather the combination of or interaction between different AI systems that lead to structural shifts in ecosystems that are responsible for adverse human rights impacts.[79]

In a similar vein, content moderation and recommendation algorithms shape the contours of cultural visibility.[80] Moderation of culturally specific expressions by AI systems may silence minority traditions or artistic forms, while AI-induced amplification of hate speech or misrepresentation of cultural groups can create a climate of hostility toward specific cultural customs.[81] Such environments can deter individuals from exercising their cultural rights, whether through restrictions on traditional dress, public

---

[74] Ditlhokwa, G. (2025). AI Representation of African Culture: Perspectives, Challenges, and Implications for Media Studies. In Artificial Intelligence and Human Agency in Education: Volume Two: AI for Equity, Well-Being, and Innovation in Teaching and Learning (pp. 175-200). Singapore: Springer Nature Singapore.

[75] United Nations. (2024). Global digital compact (A/79/L.2). https://www.un.org/global-digital-compact/sites/default/files/2024-09/Global%20Digital%20Compact%20-%20English_0.pdf

76 Christofi, A. (2023). Smart cities and cumulative effects on fundamental rights. Internet Policy Review, 12(1). Götzmann, N. (2019). Introduction to the handbook on human rights impact assessment: Principles, methods and approaches. In *Handbook on human rights impact assessment* (pp. 2-30). Edward Elgar Publishing.

[77] Joshi, P.; Santy, S.; Budhiraja, A.; Bali, K.; and Choudhury, M. 2020. The State and Fate of Linguistic Diversity and In clusion in the NLPWorld. InProceedingsofthe58thAnnual Meeting of the Association for Computational Linguistics, 6282. Association for Computational Linguistics. Bapna, A.; Caswell, I.; Kreutzer, J.; Firat, O.; van Esch, D.; Siddhant, A.; Niu, M.; Baljekar, P.; Garcia, X.; Macherey, W.; et al. 2022. Building machine translation systems for the next thousand languages. arXiv preprint arXiv:2205.03983. Benjamin, M. 2019. Empirical Evaluation of Google Translate across 107 Languages. https://www.teachyoubackwards.com/empirical-evaluation/. Accessed: 2024-07-25.

[78] Igoe, J. (2005). Becoming indigenous peoples: Difference, inequality, and the globalization of East African identity politics. African Affairs, 105(421), 399–420. DiMaggio, P. (1994). Culture and economy. The handbook of economic sociology, 27. escobal. Tahoe resources inc. (2013). www.tahoeresourcesinc.com/escobal

[79] Christofi, A. (2023). Smart cities and cumulative effects on fundamental rights. Internet Policy Review, 12(1).

[80] Dergacheva, D., & Katzenbach, C. (2024). Mandate to overblock? Understanding the impact of the European Union's Article 17 on copyright content moderation on YouTube. Policy & Internet, 16(2), 362-383.

[81] Kennedy, T. 2020, Indigenous Peoples' Experiences of Harmful Content on Social Media, Macquarie University, Sydney

religious practices, or the use of their native language, or through restrictions on participation in cultural life.[82]

From a societal perspective, the described dynamics subtly influence which forms of expression are deemed acceptable, authentic, or prestigious. The algorithmic amplification of majority cultures can marginalize minority voices and normalize cultural assimilation. Without deliberate safeguards, AI-driven cultural infrastructures risk reproducing the very inequalities that human rights law, including cultural rights, the right to development, and minority rights seek to overcome.

The structural impact of AI solutions on cultural rights and the right to development is not only confined to the equal representation of religions, political views and languages within AI models, leading to the representation or amplification of specific contents, but also touches upon the conversation on epistemic rights. Epistemic rights are broadly understood as rights relating to information, knowledge and truth.[83] AI systems, particularly generative AI, do not only convey culture and traditions, but also epistemic insights and scientific perspectives.[84] The focus of AI systems on specific views or explanations can therefore lead to a decline in the diversity of scientific, political or cultural perspectives and a skewed informational landscape. Generative AI and specifically Large Language Models (LLMs) intensify this risk: when trained predominantly on data from dominant cultures, religious contexts, scientific positions or ethical schools, they systematically underrepresent, misinterpret, or omit smaller traditions or desenting views.[85] Similar issues can also appear in the context of different historical narratives, particularly in context characterized by transitional justice, where scholarship is increasingly investigating the use of AI in the context of memorialization.[86]

A potential parthway to address such issues could be the specification of already existing human rights or other translational interpretations of existing human rights for instance the „right to access to information" that demand and require the equitable and diverse representation of opinions, views and scientific theories in AI models. Relevant normative entry points could include the UN Declaration on the Right to Development (1986), which implies epistemic diversity as a starting condition for development, or Article 15 of the International Covenant on Economic, Social and Cultural Rights, which concerns the right to benefit from scientific progress, and the right to freedom of opinion and expression. The intersection between science and culture and the aforementioned human rights context imply therefore the consideration of epistemic and cultural diversity within the design of AI solutions conveying information, particularly LLMs, but also AI systems used for educational purposes.[87]

Ultimately, the reliance of AI on extractive supply chains and its uneven ecological effects on communities in the Global Majority, but also Indigenous lands in the Global North have become increasingly controversial, particularly given that LLMs are associated with significant energy consumption that disproportionately affects regions and communities already experiencing the consequences of climate change.[88] A key example is the expansion of datacenters in water-scarce regions, intensifying conflicts over water rights due to the substantial volumes of water required for cooling these facilities.[89]

---

[82] Mohamed, S., Png, MT. & Isaac, W. Decolonial AI: Decolonial Theory as Sociotechnical Foresight in Artificial Intelligence. Philos. Technol. 33, 659–684 (2020). https://doi.org/10.1007/s13347-020-00405-8
[83] Wenar, L. (2003). Epistemic rights and legal rights. Analysis, 63(2), 142-146.
[84] Erduran, S., & Levrini, O. (2024). The impact of artificial intelligence on scientific practices: an emergent area of research for science education. International Journal of Science Education, 46(18), 1982-1989.
[85] Wright, D., Masud, S., Moore, J., Yadav, S., Antoniak, M., Christensen, P. E., ... & Augenstein, I. (2025). Epistemic Diversity and Knowledge Collapse in Large Language Models. arXiv preprint arXiv:2510.04226.
[86] Makhortykh, M., Zucker, E. M., Simon, D. J., Bultmann, D., & Ulloa, R. (2023). Shall androids dream of genocides? How generative AI can change the future of memorialization of mass atrocities. Discover Artificial Intelligence, 3(1), 28.
[87] Berendt, B., Littlejohn, A., & Blakemore, M. (2020). AI in education: Learner choice and fundamental rights. Learning, Media and Technology, 45(3), 312-324.
[88] Kaun, A., & Åker, P. (Eds.). (2023). Centering the margins of digital culture: Data centers in Sápmi, climate change denial, and the new space race. Södertörn University Library. https://www.sh.se/publications; Gordon, C. (2024, February 25). AI is accelerating the loss of our scarcest natural resource: Water. Forbes.

Taken together, the broader impacts of AI extend beyond data inputs and outputs to encompass the physical infrastructure required to sustain and expand its development, while also intersecting with cultural rights and the right to development in different normative dimensions

## 3.6 Developmental and Cultural Assumptions in the Development of AI

AI models inherently carry cultural and developmental assumptions that condition their performance, accuracy, and fairness.[90] These assumptions arise from the sociotechnical contexts in which systems are developed and deployed and are more prevalent than commonly understood.[91]

In applied contexts, such as digital health or mobility, design choices often presuppose cultural uniformity or infrastructural stability that does not exist globally. For example, telemedicine applications introduced in rural regions may rely on uninterrupted electricity or broadband connectivity, which can be unreliable.[92] When such systemic factors are not considered, the rights to health and development are indirectly affected, as certain communities are systematically excluded from digital healthcare benefits. This exclusion aligns with broader bioethical perspectives that highlight the complexities of deploying AI in medical contexts with strong cultural and developmental dimensions, thereby highlighting the need of a more contextualized interpretation of principles like autonomy and equity in the lifecycle of AI in health.[93]

The EU Horizon project MELISSA provides a clear illustration of this issue.[94] The system calculates insulin doses using visual analysis of meals for carbohydrate counting. While medically innovative, the model's performance varies depending on cultural eating habits, such as fasting during Ramadan or vegetarianism among specific populations.[95] These factors influence dietary patterns and, if not

---

https://www.forbes.com/sites/cindygordon/2024/02/25/ai-is-accelerating-the-loss-of-our-scarcest-natural-resource-water/

[89] Siddik, M. A. B., Shehabi, A., & Marston, L. (2021). The environmental footprint of data centers in the United States. Environmental Research Letters, 16(6), 064017.

[90] Azizov, D. (2024). From Idioms to Algorithms: Translating Culture-Specific Expressions in AI Systems. Iconic Research And Engineering Journals, 7(10), 543-551. Fletcher, R. R., Nakeshimana, A., & Olubeko, O. (2021). Addressing fairness, bias, and appropriate use of artificial intelligence and machine learning in global health. Frontiers in artificial intelligence, 3, 561802.

[91] Szykman, G. A., Brandão, A. L., & Gois, J. P. (2018). Development of a gesture-based game applying participatory design to reflect values of manual wheelchair users. *International Journal of Computer Games Technology, 2018*, Article 2607618. https://doi.org/10.1155/2018/2607618

[92] Farhat, R., Malik, A. R. A., Sheikh, A. H., & Fatima, A. N. (2024). The role of AI in enhancing healthcare access and service quality in resource-limited settings. International Journal of Artificial Intelligence, 11(2), 70-79. Barry, R., Green, E., Robson, K. et al. Factors critical for the successful delivery of telehealth to rural populations: a descriptive qualitative study. BMC Health Serv Res 24, 908 (2024). https://doi.org/10.1186/s12913-024-11233-3

[93] Amugongo, L. M., Kriebitz, A., Boch, A., & Lütge, C. (2025). Operationalising AI ethics through the agile software development lifecycle: a case study of AI-enabled mobile health applications. AI and Ethics, 5(1), 227-244.

[94] MELISSA Consortium. (2022). Mobile artificial intelligence solution for diabetes adapted care (MELISSA) (EU Horizon Project No. 101057730). https://www.melissa-diabetes.eu/

[95] Panagiotou, M., Strømmen, K., Brigato, L., de Galan, B. E., & Mougiakakou, S. (2025). Role of artificial intelligence in enhancing insulin recommendations and therapy outcomes. Die Diabetologie, 1-8.

integrated into the training process, reduce accuracy and fairness. AI systems must therefore include cultural awareness as performance and safety parameters in healthcare contexts.[96]

Similarly, cultural assumptions surface in the development of autonomous driving systems. Models trained primarily in European or North American environments tend to assume predictable traffic behaviour and strict compliance with traffic laws.[97] When applied in regions with different road dynamics, traffic densities, or social conventions—such as informal pedestrian crossing patterns—these systems may misinterpret behaviour or increase accident risk. These examples underscore the need for adaptive, context-sensitive training datasets that reflect the cultural and developmental realities of local infrastructure.

Similar concerns also appear in the context of education.[98] Education in Sub-Saharan Africa places a strong cultural emphasis on social interpersonal skills.[99] However, many educational technologies are built around the values of individualized learning. Tools like ChatGPT similarly promote individualized, screen-based learning, which undermines a culture of communal, collaborative, and interpersonal learning in sub-Saharan Africa.[100]

Neglecting these realities constitutes not only a technical limitation but also a human rights issue. The resulting biases and exclusion patterns disproportionately affect individuals and groups whose behaviour, language, or living conditions fall outside the assumed norm.[101] Such structural insensitivity leads to indirect discrimination and impedes the equal enjoyment of specific rights, including potential or actual adverse impacts on the right to physical integrity, the right to health or the right to education.

Biases are therefore not only embedded in AI algorithms; they may originate in the policy frameworks that govern them. When regulatory definitions of "high risk" AI systems are overly narrow or universalist, they may fail to capture cultural vulnerabilities and developmental disparities. Legal regimes must guard against reinforcing the dominance of industrialized cultural norms and technological infrastructures, which could perpetuate epistemic and socio-cultural inequalities.

## 3.7 The Need for a Comprehensive Perspective

The preceding analysis demonstrates the intrinsic relationship between AI, cultural rights, and the right to development. Both cultural rights and the right to development are not only expressions of human dignity and human flourishing but also essential conditions enabling these wider considerations. While they are intertwined with other human rights, they have distinct meanings and require their own attention and human rights instruments. In addition, the development of AI itself carries important developmental and cultural dimensions, as it increasingly influences the evolution of culture and socio-

---

[96] Jain, A., Brooks, J. R., Alford, C. C., Chang, C. S., Mueller, N. M., Umscheid, C. A., & Bierman, A. S. (2023). Awareness of racial and ethnic bias and potential solutions to address bias with use of health care algorithms. JAMA Health Forum 4 (6), e231197–e231197 (2023).

[97] Shah, K., & Guven, E. (2025, March). Exploring Ethical Issues and Challenges of Autonomous Vehicles in Different Countries and Cultures. In 2025 59th Annual Conference on Information Sciences and Systems (CISS) (pp. 1-6). IEEE. Serrano, S. M., Méndez Blanco, O., Worrall, S., Sotelo, M. A., & Femández-Llorca, D. (2024). Cross-cultural analysis of pedestrian group behaviour influence on crossing decisions in interactions with autonomous vehicles. In 2024 IEEE 27th International Conference on Intelligent Transportation Systems (ITSC) (pp. 2006–2012). IEEE. https://doi.org/10.1109/ITSC58415.2024.10919759

[98] Majgaard, K., & Mingat, A. (2012). Education in sub-Saharan Africa: A comparative analysis. World Bank Publications.

[99] Maisiri, J., & Musonza, S. (2024). The cultural cost of AI in Africa's education systems. UNESCO. https://www.unesco.org/en/articles/cultural-cost-ai-africas-education-systems

[100] Majumder, M., & Tripathi, A. K. (2020). Transformative power of technologies: Cultural transfer and globalization. AI & Society, 38, 2295–2303. https://doi.org/10.1007/s00146-021-01144-w. van de Poel, I., & Kroes, P. (2014). Can Technology Embody Values? In P. Kroes & P.-P. Verbeek (Eds.), The Moral Status of Technical Artefacts (pp. 103–124). Springer Netherlands. https://doi.org/10.1007/978-94-007-7914-3_7.

[101] Chen, Z. Ethics and discrimination in artificial intelligence-enabled recruitment practices. Humanit Soc Sci Commun 10, 567 (2023). https://doi.org/10.1057/s41599-023-02079-x

Cultural Rights and the Rights to Development in the Age of AI:
Implications for Global Human Rights Governanceeconomic developments. The right to development and cultural rights are therefore important lenses to evaluate the global, regional and national impacts of AI on human flourishing.

Breaking down the impact of AI on cultural rights and the right to development into distinct categories is essential to fully grasp the multidimensional role of the nexus between cultural rights, the right to development, and AI.

Most importantly, AI development itself is an expression of human creativity, which is protected by international human rights law through general provisions on human autonomy, scientific and intellectual freedom. However, due to inherent epistemic differences between humans and machines, particularly the statistical nature and data dependency of AI, AI systems lack genuine creativity. This distinction can result in different types of impacts on human rights linked to or caused by AI—not only through its outputs but also through its inputs and its context-sensitive treatment of individual and collective rights. Particularly, the distinction between impacts linked to or caused by AI is relevant to differentiate between first-order and second-order impacts of AI on rights. Additionally, cultural and developmental contexts are crucial parameters that are helpful to locate but also attribute AI-borne human rights impacts.[102] All these impacts are further amplified by the complexity and resulting diffusion of algorithmic accountability across global AI value chains, which involve diverse actors responsible for data provision, hypothesis formulation, and variable selection. While it is unlikely that AI will engage in a perfect representation of existing cultures, languages, traditions and ultimately scientific perspective, the only meaningful way is to shape the further development of AI in a sense that comes increasingly closer to these more nuanced understandings and the alignment with relevant cultural and developmental rights, but also epistemic rights.

Apart from this necessary conceptualization, existing discussions on the right to development and cultural rights, most importantly the capabilities approach, are critical for resolving normative conflicts arising from different perspectives on the topic, for instance, the balancing act between innovation and cultural rights in low-resource settings. For example, the meaningful participation of the Global South in the AI race and the acceptance of variations in implementing common international AI norms and standards are essential, given developmental and infrastructural differences including access to electricity, (digital) literacy levels, and other existing divides.

If global governance fails to recognize the collective dimensions of cultural and developmental rights, it risks reinforcing existing hierarchies instead of addressing them. The right to development and cultural rights, as part of international human rights law, are therefore essential considerations to prevent the emergence of a biased model for global AI governance.

## 4 Cultural Rights and The Right to Development in Emerging AI Governance

Increasing recognition of human rights concerns in AI governance is evident across international, regional, and scholarly platforms.[103] Notably, the United Nations and affiliated agencies, such as UNESCO and the Office of the High Commissioner for Human Rights (OHCHR), have adopted human rights-based approaches in recommendations, resolutions, and consultative frameworks.[104] The UN Digital Global Compact and recent UN General Assembly resolutions such as A/78/L.49 demonstrate a growing global consensus on the need to address the impact of AI on human rights, including cultural rights and the right to development.

---

[102] This connects to earlier attemps to categorize between different causal connections between algorithmic impacts and human rights. See: Kriebitz, A., & Lütge, C. (2020). Artificial intelligence and human rights: a business ethical assessment. Business and Human Rights Journal, 5(1), 84-104.

[103] Hogan, L., & Lasek-Markey, M. (2024). Towards a Human Rights-Based Approach to Ethical AI Governance in Europe. Philosophies, 9(6), 181. Bakiner, O. (2023). The promises and challenges of addressing artificial intelligence with human rights. Big Data & Society, 10(2), 20539517231205476.

[104] Kriebitz, A. (2024). Protecting and realizing human rights in the context of artificial intelligence: a problem statement. In The Elgar Companion to Applied AI Ethics (pp. 95-108). Edward Elgar Publishing.

Despite these advances, the international landscape remains fragmented. There is no binding, globally applicable AI regulatory framework based on international law in sight that provides comprehensive protection for cultural rights or the right to development. Instead, governance is shaped by a variety of non-binding declarations, regional conventions, national legislations, and evolving scholarly proposals.

This situation raises an important question about the extent to which cultural rights and the right to development are integrated into existing global, regional, and national AI governance approaches.

### 4.1. Non-Binding Frameworks

Global deliberations on AI and human rights have largely coalesced around non-binding recommendations and best practices. The UNESCO Recommendation on the Ethics of Artificial Intelligence (2021) is a landmark document that emphasizes human rights, human dignity, cultural diversity, and international humanitarian law as foundational pillars of ethical AI governance. While influential, the Recommendation is not enforceable on its own and instead serves as guidance for policy development and capacity building in member states.[105] UNESCO has also advanced AI literacy initiatives, stressing the importance of education as a tool to empower culturally diverse societies in the digital age.

Alongside the United Nations' deliberations, the 38 member countries, the European Union, and 11 non-member countries endorsed the OECD AI Principles.[106] Although not framed as a cultural rights instrument, the ethical commitments are central to the realization of cultural rights and the right to development. The first principle, *inclusive growth, sustainable development, and well-being*, requires that AI systems reduce social and economic inequalities and include underrepresented populations. The second principle, *human-centred values and fairness*, is central to cultural rights, which protect vulnerable communities from assimilation pressures and require equal treatment for different languages and ways of life. The third principle, *transparency and explainability*, allows communities to understand which data AI systems use to represent their cultures, which is essential to ensuring their identities and histories are presented accurately. The fourth principle, *robustness, security, and safety*, is essential given the vast history of human rights abuses that point to ways bad actors have used information to target vulnerable communities, caricaturize viewpoints and histories, and manipulate public discourse about cultural and religious minorities. The fifth principle, *accountability,* ensures that individuals and groups have the right to seek an effective remedy should AI systems undermine their cultural rights, such as exploiting artistic expression without consent, eliminating or undermining minority languages from AI systems, or amplifying harmful speech and misinformation. Together, these principles illustrate the growing demand for countries to ensure that AI protects and advances cultural rights and the right to development.

Similar discussions within the International Telecommunication Union (ITU)—notably the "AI for Good" program—address sustainable development and digital equity, encouraging the responsible use of AI to advance the right to development.[107] Efforts are made to ensure that AI benefits vulnerable populations, including minorities and indigenous communities, though institutional enforcement and monitoring remain limited.

---

[105] United Nations Educational, Scientific and Cultural Organization (UNESCO). (2021). Recommendation on the ethics of artificial intelligence. Retrieved from https://unesdoc.unesco.org/ark:/48223/pf0000379985;
[106] Organisation for Economic Co-operation and Development. "Recommendation of the Council on Artificial Intelligence (OECD AI Principles), OECD Legal Instruments, adopted May 22, 2019, amended May 3, 2024, https://legalinstruments.oecd.org/en/instruments/OECD-LEGAL-0449
[107] Walshe, R., Casey, K., Kernan, J., & Fitzpatrick, D. (2020). AI and big data standardization: Contributing to United Nations sustainable development goals. Journal of ICT Standardization, 8(2), 77-106.

The Global Digital Compact and related initiatives by the BRICS countries and the African Union increasingly foreground the right to development as both a principle and a policy objective.[108] These instruments reflect diverging priorities, with the Global Majority (especially Africa, Latin America and parts of Asia) advocating strongly for technology-driven development alongside protection of cultural identity and heritage. Such divergence underscores differing global approaches regarding the balance between innovation and rights protection.

The UN Guiding Principles on Business and Human Rights (UNGPs) and sectoral publications by the UN Global Compact Network have introduced the concept of corporate responsibility for AI impacts, including company obligations to respect cultural rights and the interests of indigenous groups.[109] These frameworks refer explicitly to the International Covenant on Economic, Social, and Cultural Rights (ICESCR), as well as the emerging "right to a clean and healthy environment," yet remain reliant on voluntary compliance with limited sanction mechanisms for violations. They illustrate the importance of AI companies and businesses that deploy AI in their operations to take responsibility for their human rights due diligence obligations.

### 4.2. Regional and Supranational Approaches

Regional conventions vary significantly in how they recognize and protect cultural rights and the right to development in the AI context. The Council of Europe Framework Convention on AI, Human Rights, Democracy and Rule of Law is a pioneering multilateral treaty rooted in the European human rights system, yet it is also open to non-European states such as Japan and the United States.[110] While it establishes important principles for human rights-based AI governance, it notably lacks explicit recognition of collective rights, which are crucial for effectively safeguarding minority communities that are protected by cultural rights, but also provisions addressing developmental considerations. This aligns with a wider tendency among Western states that see the codification of the right to development as critical. This is relevant against the backdrop of the lack of constitutional consensus in States like the United States on the extent of the progressive realization of socio-economic rights, even within their own territory, let alone in the global context. Furthermore, as Convention 108+ on data protection, the Framework Convention does not address corporate responsibility and primarily focuses on state obligations to enact laws and policies protecting human rights within their digital domains. Similar to the EU AI Act, this approach follows a risk-based approach to prevent adverse human rights impacts created by AI. This framework has shown gaps in civil society involvement and representation from the Global South, especially African stakeholders.[111] Data and input-related issues fall outside its scope, as they are covered by the broader European human rights system under Convention 108+ and accompanying legislation such as the GDPR and the UK GDPR.

Recent developments in Europe and the Americas reveal attempts to roll back existing human rights advancements in AI governance, exemplified by U.S. executive orders issued in the second term of President Trump in 2025 that signal a weakening of established standards.[112] Despite the United States

---

[108] BRICS Policy Center. (2025). Artificial Intelligence governance in BRICS: Cooperation and development for social inclusion. https://brics.br/en/news/articles/artificial-intelligence-governance-in-brics-cooperation-and-development-for-social-inclusion

[109] UN Global Compact Network Germany. (2024). Artificial intelligence and human rights: Recommendations for companies. https://www.globalcompact.de/fileadmin/user_upload/-Dokumente_PDFs/artificial_intelligence_and_human_rights_EN.pdf

[110] Council of Europe (CoE). (2024). Framework Convention on Artificial Intelligence and Human Rights, Democracy, and the Rule of Law. Retrieved from https://www.coe.int/en/web/artificial-intelligence;

[111] See: (Kriebitz, A., & Corrigan, C. C. (Hrsg.). (2025). *Promoting and advancing human rights in global AI ecosystems: The need for a comprehensive framework under international law*. Ethics Lab at Rutgers University. https://aiethicslab.rutgers.edu/publications/promoting-and-advancing-human-rights-in-global-ai-ecosystems/

[112] Mackowski, M. J., Maschek, W. A., Goldstein, B. L., Jacobson, J. B., Friel, A. L., & Kirk, M. (2025, February 10). Trump's AI executive order: A shift toward deregulation. The National Law

being a signatory, the Framework Convention has not prevented this regression, underscoring its limitations in enforcing binding commitments, particularly with states hesitant to adopt legislative protections for human rights.

In contrast, the African human rights tradition aligns more closely with the right to development and minority protection.[113] Instruments such as the African Charter on the Rights and Welfare of the Child and, notably, the Charter for African Cultural Renaissance emphasize concepts like "cultural democracy" that resonate profoundly in AI governance debates, particularly in safeguarding Africa's rich cultural diversity and endangered languages suppressed during colonialism. Ongoing discussions about an African-specific AI and human rights convention reflect these unique local realities, highlighting the continent's aspiration for equitable access to technology and culturally sensitive governance, as well as a stronger pivot toward embedding AI governance within both cultural rights and the right to development. In this sense, the conversation in the African context seems to come closest in addressing the aforementioned issues of epistemic rights, as well as the need to equitable represent different views in culturally relevant ecosystems and epistemic communities, considering both the individual rights angle, but also notion of collective cultural and developmental rights.

## 5. The Integration of Cultural Rights and the Right to Development in Emerging Approaches

The divergence in international governance of AI is not a coincidence but reflects different regional challenges, political preferences, and historical experiences that are relevant to the relative status of cultural rights and the right to development within AI governance. Apart from the UNESCO Recommendation on the Ethics of Artificial Intelligence, existing binding instruments show relatively limited integration of development and cultural human rights considerations.

A key issue resides in the omission of identifying the duties of various actors, such as developers of large language models, social media platforms, and other AI systems regarding their responsibility to address structural impacts on cultural rights in both their collective and individual dimensions. Additionally, the international community has made little progress in defining the implications of the right to development for state obligations within the context of AI. Equally important is establishing clear parameters to determine when an AI system reaches a critical mass from cultural and developmental perspectives, guiding appropriate regulatory and ethical responses.

In the following section, we take a more detailed look on the relevant processes of integrating cultural rights and the right to development within existing regional and international conversations on AI governance.

### 5.1. Asymmetric Protection of Cultural Rights and the Right to Development in AI Governance

Despite an expanding body of non-binding documents and regional instruments, meaningful enforcement and harmonization across different jurisdictions remain difficult to achieve without globally binding rules. This owes to the growing contestation of the right to development and cultural rights, but also to a growing divergence of AI governance.

Enforcement of minority protections and cultural rights varies significantly, reflecting differing national priorities, historical narratives and geopolitical challenges.[114] Moreover, growing societal polarization including many Western societies poses increasing risks of eroding cultural rights. External political conflicts exacerbate divides through phenomena such as Islamophobia, Antiziganism, and

---

Review. https://natlawreview.com/article/key-insights-president-trumps-new-ai-executive-order-and-policy-regulatory

[113] Beall, K. M. (2022). The Global South and global human rights: international responsibility for the right to development. *Third World Quarterly*, *43*(10), 2337-2356.

[114] Symonides, J. (1998). Cultural rights: a neglected category of human rights. International Social Science Journal, 50(158). Kalaycioglu E, Subotić J. Global politics of cultural heritage: Status, authority, and geopolitics. *Review of International Studies*. Published online 2025:1-13. doi:10.1017/S0260210525101502

Antisemitism, challenging the protection and realization of cultural and minority rights.[115] Furthermore, the increasing use of deep fakes and the circulation of ethnic minority caricatures particularly amplify these risks.[116] Such developments underscore the non-negotiable human rights obligations of states to protect cultural rights domestically, with particular vigilance to prevent irreversible human rights harms.[117]

Similar challenges in enforcement are also found in respect to the right to development. In developing countries, challenges extend beyond infrastructural deficits such as unreliable electricity, limited digital literacy, and inadequate remedial mechanisms to the often poor fit of legislative frameworks designed for digitally advanced societies. Early legislation focused mainly on constitutionally guaranteed rights, predominantly individual rights as understood in many Global North legal systems, emphasizing prohibitions and specific safeguards for AI development and deployment.[118] When such frameworks are applied in developing countries without sufficient flexibility for different developmental contexts, they risk stifling innovation or the implementation of the right to development. There is thus a critical need for states to exercise discretion by factoring in developmental considerations and the progressive realization of socio-economic rights within a wider response to AI governance on a global level.

This divergence highlights deeply rooted differences in global AI governance priorities. The European Union, OECD and the Council of Europe emphasize individual fundamental rights rooted in post-World War II European human rights traditions—personal freedoms, autonomy, privacy, and non-discrimination in digital environments. The United States, meanwhile, has largely disengaged overarching federal regulation, favouring voluntary guidelines and sector-specific measures that underscore market-led innovation over binding rules, but also the right of individuals to engage in AI innovation. China pursues a people-centered AI governance model that positions the technology as a driver of national competitiveness, social stability, and modernization with a stronger focus on collective interests.[119] Finally, the Global Majority, including Africa, Latin America, and parts of Asia, seem to prioritize collective rights, cultural self-determination, and the protection of historically marginalized communities. These regions foreground the legacies of colonialism, cultural suppression, and assimilation pressures, placing the right to development, equitable access to AI infrastructures and cultural rights at the centre of their normative order. They express concern that the dominance of a few Global North states in setting AI standards may amount to digital colonialism, threatening cultural autonomy and developmental sovereignty.

Taken together, the normative differences on the integration of cultural and development rights do not only connect to overarching shifts in international law, but also to already existing divergencies in normative preferences, when it comes to balancing different human rights aspects. Important dimensionalities are here different perspectives on individual and collective rights, the status of minorities within international human rights law, and the derogation between technological openness and stricter enforcement of human rights protections. These are embedded themselves in different cultural and developmental context and intersect with the interests of UN member states, but also regional organizations which are characterized by compatible similar living standards and moral traditions.

---

[115] Taras, R. (2015). 'Islamophobia never stands still': race, religion, and culture. In Racialization and Religion (pp. 33-49). Routledge.
[116] Scheiber, M. (2024). Antisemitische Deepfakes: Dekonstruktion über Bildwissen. Politikum, 10(4), 46-51.
[117] United Nations General Assembly. (2025). Human Rights and Artificial Intelligence (A/80/287). United Nations. https://docs.un.org/en/A/80/287
[118] EU AI Act: Regulation (EU) 2024/1689 of the European Parliament and of the Council of 13 June 2024 laying down harmonised rules on artificial intelligence. Official Journal of the European Union. (14.7.2024). Regulation (EU) 2024/1689 of the European Parliament and of the Council. (2024). https://eur-lex.europa.eu/eli/reg/2024/1689/oj/eng
[119] Global Times. (2024, November). China's 'people-centered, AI-for-good' principles shine at WIC Wuzhen Summit. https://www.globaltimes.cn/page/202411/1323422.shtml
WIPO China. (2024). Embracing a people-centered and AI-for-good digital future. https://www.wipo.int/en/web/office-china/w/news/2024/wipo-china-embracing-a-people-centered-and-ai-for-good-digital-future

## 5.2. Emerging Integrative Approaches

Apart from the emergence of different regional perspectives on AI governance, the past months have witnessed an intensifying conversation on global rules. Existing policy dialogues and collaborative efforts such as the *North-South Policy Exchange on a UN Convention on AI, Data and Human Rights* in Stellenbosch, South Africa, and the *Munich Convention on Artificial Intelligence, Data and Human Rights* draft illustrate attempts to harmonize diverse traditions and establish unified normative foundations addressing these divergent priorities.[120] More recently, the UN has itself engaged in a global discussion on AI governance and several initiatives have started to address global red lines for AI development.

The Munich Convention, initiated in July 2024 by international and interdisciplinary scholars, builds on established international law and carefully incorporates various international understandings and interpretations of human rights. Notably, it integrates perspectives from both the Global Majority and Global North, aiming to develop a framework grounded in United Nations principles while fostering interoperability between emerging European and African human rights traditions concerning AI and digital technologies. For example, Article XII ("Principles for Artificial Intelligence Systems Used in an International Context") requires states to consider political, cultural, and developmental contexts in cross-border AI deployment and mandates the participatory, fair, and consented representation of indigenous peoples and ethnic minorities. The Convention's preamble and substantive provisions explicitly link AI governance to sustainable development objectives and stress the need for context-sensitive, equitable regulation to prevent the perpetuation of digital inequalities and to protect vulnerable groups.

Complementing this, the UNESCO Recommendations on the Ethics of AI provide a comprehensive basis for international AI regulation by highlighting the integration of cultural and developmental rights into responsible AI governance. Embedding these rights at the United Nations level is critical, especially as domestic political shifts can erode constitutionally guaranteed rights. Care must be taken to ensure that AI regulation itself does not become a repressive instrument, particularly in states where cultural identity is closely tied to national identity and dominant value systems prevail. This concern grows in importance alongside the general reduction of funding, threatening the right to development.

Strengthening cultural rights and the right to development might be essential, not only from the perspective of a rights-based approach to global AI governance but also from the perspective of reaffirming these rights in the digital context. The current erosion of these dimensions risks significant implications under customary international law, and their absence in any international legal framework addressing AI and other sociotechnical systems could likely have profound negative consequences on the relative status of these rights within international human rights law, contrary to the principles of the United Nations.

## 5.3 The Role of Corporate Human Rights Responsibilities in the AI, Culture and Development Nexus

Owing to the critical role of multinational enterprises within the AI lifecycle, corporate human rights responsibilities are essential for preventing structural adverse impacts on cultural rights and the right to development.[121] Frameworks such as the UNGPon Business and Human Rightsform an important

---

[120] Institute for Ethics in Artificial Intelligence. (2025, September 7). From Munich to Stellenbosch: Building momentum for a global AI and human rights framework. Technical University of Munich. https://www.ieai.sot.tum.de/building-momentum-for-a-global-ai-and-human-rights-framework/

[121] Srivastava, S., & Bullock, J. (2024). AI, global governance, and digital sovereignty. arXiv preprint arXiv:2410.17481. UNESCO. (2021). Recommendation on the Ethics of Artificial Intelligence. Paris: UNESCO. https://unesdoc.unesco.org/ark:/48223/pf0000377897 Shaleen Khanal, Hongzhou Zhang, Araz

foundation, given their increasing relevance in AI governance discussions.[122] However, researchers highlight that the UNGPs lack the specificity and translational guidance needed for direct application to the challenges posed by AI.[123]

Accordingly, the integration of AI within existing concepts of corporate human rights responsibilites can draw from the UNESCO Recommendations on Ethics of AI, as they already encourage companies to address developmental and cultural contexts, as well as linguistic inclusion and minority representation in particular, in their AI operations. Another important vehicle for reinforcing corporate responsibilities is the explicit inclusion of AI in already existing practices such as algorithmic impact assessments.[124] The Munich Convention initiative exemplifies this by referencing "Principles for Artificial Intelligence Systems Used in an International Context" and emphasizing state action to address political, cultural, and developmental factors in cross-border deployments, by leveraging the UNGPs as an important entry point.[125]

Ultimately, the success of these measures hinges not only on multi-stakeholder collaboration encompassing governments, civil society, and local communities[126], but also the leadership and willingness of UN member states to prevent practices such as "AI washing," where organizations adopt ethical principles or sustainability language without implementing substantive changes in their AI systems or business models.[127] Hence, the contextualization of human rights responsibilities to the AI within the AI-culture-development nexus might require an underlying binding framework, which reinforces the extraterritorial human rights protections and specifies the meaning of these in the AI context.

---

Taeihagh, Why and how is the power of Big Tech increasing in the policy process? The case of generative AI, Policy and Society, Volume 44, Issue 1, January 2025, Pages 52–69, https://doi.org/10.1093/polsoc/puae012

[122] UN General Assembly (2024). Resolution A 78.L49. Retrieved from: https://docs.un.org/en/A/78/L.49

[123] Kriebitz, A., & Corrigan, C. C. (Eds.). (2025). *Promoting and advancing human rights in global AI ecosystems: The need for a comprehensive framework under international law*. Ethics Lab at Rutgers University. https://aiethicslab.rutgers.edu/publications/promoting-and-advancing-human-rights-in-global-ai-ecosystems/

[124] Schertel, L., & Stray, J. (2024). AI as a Public Good: Ensuring Democratic Control of AI in the Information Space. In Forum on Information & Democracy. February.

[125] Kriebitz, A., Corrigan, C. C., & Boch, A. (2024, October). Munich Convention on Artificial Intelligence, Data and Human Rights (Draft for Public Consultation). Presented at the International Summit on Artificial Intelligence and Human Rights, Munich. Institute for Ethics in Artificial Intelligence. https://www.researchgate.net/publication/384678262_Munich_Convention_on_Artificial_Intelligence_Data_and_Human_Rights_Draft_for_Public_Consultation

[126] OECD, F. T. C. A. N. (2019). Principles on Artificial Intelligence.

[127] Floridi, L., Cowls, J., Beltrametti, M., et al. (2018). AI4People—An ethical framework for a good AI society: Opportunities, risks, principles, and recommendations. Minds and Machines, 28(4), 689–707. https://doi.org/10.1007/s11023-018-9482-5. Gantzias, G. (2020). Dynamics of public interest in artificial intelligence:'Business intelligence culture'and global regulation in the digital era. In The Palgrave handbook of corporate sustainability in the digital era (pp. 259-281). Cham: Springer International Publishing.

# 6. Conclusion

Cultural rights and the right to development have emerged as important cornerstones of international human rights law. They are not only instrumental for the understanding of related rights, including freedom of expression, but also norms outside of traditional human rights law, including the right to self-determination and the rights of ethnic minorities. Hence, cultural rights and the right to development carry important individual and collective connotations and contribute an interconnected normative space aligned with UN Charter and the Universal Declaration of Human Rights.

The introduction of AI within the space described and protected by both rights raises fundamental questions. The development, deployment, and use of AI can be understood as expressions of human creativity, which is itself protected by cultural and economic rights such as artistic freedom, freedom of opinion and expression, and entrepreneurial freedom. Nevertheless, if the development of AI proceeds without appropriate oversight and governance, it risks eroding the protection of cultural rights an the right to development. Scholarship has identified several areas, where AI is already affecting cultural diversity, leading to the marginalization of minority languages and reinforcing cultural hegemony. Moreover, cultural assumptions are also relevant factors in the design of AI systems and failure to acknowledge this can introduce safety and health risks. In addition, the development and dissemination of AI take place in a context characterized by unequal starting positions, in part resulting from earlier human rights violations, such as colonialism. As a result, many stakeholders, particularly from marginalized communities, have warned that AI might pose structural challenges to the realization of the right to development and cultural rights in global contexts. The impact of these developments extends beyond specific rights to affect epistemic diversity and the dissemination of knowledge within society. The use of AI in cultural and developmental contexts therefore engages multiple dimensions of human rights and calls for a careful consideration of the rights at stake, which are deeply connected to fundamental anthropological and societal questions.

Thus far, existing protections under national, regional, and emerging human rights instruments remain inadequate to address the complex and multifaceted challenges posed by AI. AI regulation has been mainly focusing on constitutionally enshrined rights of UN member states, such as human dignity, the right to non-discrimination, the right to privacy or the right to remedy. While the focus on constitutional rights is warranted, it does not, in itself, meaningfully protect cultural rights and the right to development in the international human rights legal space. The cultural and developmental dimension of AI has a strong extraterritorial effect. Additionally, the current situation risks embedding strong cultural and developmental biases in AI governance, as these dimensions are largely absent from binding frameworks. This is concerning because cultural and developmental assumptions critically influence other human rights. Failure to address these issues may exacerbate governance asymmetries and cause tangible harms, particularly between the Global North and South, including adverse impacts on physical integrity and the right to health due to biased decision-making in critical stages of the AI lifecycle.

As a result, the transformative impact of AI on cultural expression—from artistic creation to the representation of languages and minority identities—as well as on equitable development opportunities, requires a binding international instrument as an interpretative tool that translates existing human rights instruments, for instance corporate human rights responsibilities, into the emerging context of AI. The human rights impact of AI is inherently international, given the history of international law as a vehicle to protect the rights of minorities across jurisdictions and as an instrument to protect the right to development, particularly on a global scale, meaningfully. In addition, cultural rights and the right to development are themselves increasingly challenged, particularly when it comes to cuts in international aid or the growing persecution of ethnic, religious or cultural minorities. Safeguards at the United Nations level are therefore increasingly important to prevent states from failing to adhere to existing obligations under international human rights law in the AI context.

While the paper is a preliminary assessment of the cultural and developmental impact of AI, it identifies further avenues for policy making and scholarship. The current patchwork of legislation, such as the EU AI Act and regional treaties like the Council of Europe Framework Convention, exhibits notable gaps, especially in recognizing the right to development and the meaningful protection of cultural and linguistic minorities. The reaffirmation of these rights within a binding instrument on AI and human rights is needed in order to create a comprehensive rights-based approach to AI governance on global level and to reaffirming cultural rights and the right to development in an increasingly challenging international environment. To further substantiate this claim, future research needs to further investigate the cultural and developmental dimension of AI and AI regulation, as well as behavioral impact of cultural and developmental biases on decision-making in the AI lifecycle as a necessary safeguard for cultural plurality, epistemic diversity, and equitable participation in the digital future.